\documentstyle[aps,prl,twocolumn]{revtex}

\begin{document}
\author{V. V. Dobrovitski, M. I. Katsnelson\cite{mik}, and B. N. Harmon}
\address{Ames Laboratory, Iowa State University, Ames, Iowa, 50011}

\title{Mechanisms of decoherence in weakly anisotropic molecular magnets}

\draft

\maketitle

\begin{abstract}
Decoherence mechanisms in crystals of weakly anisotropic magnetic molecules,
such as V$_{15}$, are studied. We show that an important decohering factor
is the 
rapid thermal fluctuation of dipolar interactions between magnetic
molecules. A model is proposed
to describe the influence of this source of decoherence.
Based on the exact solution of this model, 
we show that at relatively high temperatures, about 0.5 K, the 
quantum coherence in a V$_{15}$ molecule is not suppressed, and, in 
principle, can be detected experimentally. Therefore, these 
molecules may be suitable prototype systems for study of physical processes
taking place in quantum computers.
\end{abstract}


A new class of magnetic compounds, molecular magnets \cite{molmag}, has 
been attracting much attention. Each molecule of such a compound 
is a nanomagnetic entity with a large spin (or, in the antiferromagnet
case, large staggered magnetization). The interaction 
between different molecules, being of the dipole-dipole type, is very
small, so that the corresponding crystal is an arrangement of identical
weakly interacting nanomagnets. 
Molecular magnets are ideal objects to
study phenomena of great scientific
importance for mesoscopic physics, such as spin 
relaxation in nanomagnets, quantum tunneling of magnetization, 
topological quantum phase interference, quantum coherence, etc
\cite{qtm}.  Low-spin weakly
anisotropic compounds, like V$_{15}$ \cite{V15,butterfly}, 
demonstrate well-pronounced quantum properties, such as significant
tunneling splitting of low-lying spin states. As we show here,
they are attractive
prototype systems to study mesoscopic quantum coherence
and physical processes which destroy it. Besides fundamental
science, these studies are important also for the 
implementation of quantum computation \cite{qucomp}.

At present, for strongly anisotropic high-spin magnetic molecules such as
Mn$_{12}$ and Fe$_8$, different kinds of decohering
interactions have
been studied \cite{dissip,stamp} and their
interplay with quantum properties at low temperatures (below 1.5--2 K)
is well understood. A general conclusion \cite{stamp} about 
strongly anisotropic
systems is that the dissipative environment, especially the
bath of nuclear spins, rapidly destroys coherence even at
very low temperatures, and only incoherent tunneling survives.

Decoherence in weakly anisotropic magnetic molecules 
has not yet received much study, and such a study is
the main purpose of the present paper. We analyze various sources of 
decoherence for such molecular magnets as V$_{15}$, and
show that
in the temperature range 0.2 K -- 0.5 K the decoherence
is governed by rapidly fluctuating dipole-dipole fields
produced by thermally activated molecules.
This mechanism in molecular magnets has not been 
considered before; estimates show that in strongly anisotropic magnets
like Mn$_{12}$ or Fe$_8$ this effect is small.
Based on an exactly solvable model, we demonstrate that even
at temperatures as high as 0.5 K, the quantum coherence 
in V$_{15}$ molecules is remarkably robust, and, in principle, can
be detected experimentally. Therefore, the V$_{15}$ molecular magnet is
a promising candidate for the study of quantum coherence and may
be a useful prototype system for the investigation of physical 
processes taking place in quantum computers.

The magnetic subsystem of the molecule 
K$_6$[V$_{15}$As$_6$O$_{42}$(H$_2$O)]$\cdot$8H$_2$O, 
(denoted for brevity as V$_{15}$) consists of fifteen V$^{4+}$ 
ions with the spin 1/2 (see Fig.\ \ref{v15scheme}).
The ions form two nonplanar hexagons (with total spin equal
to zero) and a triangle sandwiched between them. 
Exchange interactions between  
ions are reasonably large (from 30 K to 800 K), but, due to the strong
spin frustration present in the molecule, the couplings
of the central triangle spins with the hexagons cancel
each other (see Fig.\ \ref{v15scheme}). 
The hexagon spins form a rather stiff antiferromagnetic
structure, and the low-energy part of the spectrum is
defined by only three weakly coupled spin 1/2 ions belonging to
the central triangle. An effective
exchange coupling between the triangle spins $J\simeq 2.5$ K 
is present. Thus, the ground-state term consisting of two doublets 
with $S=1/2$ is separated from the low-lying excited term $S=3/2$
by the distance $\Delta_1 = 3J/2 \simeq 3.8$ K.
Experimental results \cite{butterfly} suggest that
within the two ground-state doublets, 
the states $|S=1/2,S_z=+1/2\rangle$ and $|S=1/2,S_z=-1/2\rangle$
are mixed (a small anisotropic interaction may be responsible,
but it is not of concern for the arguments presented), so
that tunneling between these levels occurs and 
the fourfold degeneracy of the ground state is partly lifted
(it cannot be
lifted completely because of Kramers' theorem: in the absence of
an external field all levels are doubly degenerate). 
The coherent tunneling leads to a splitting
$\Delta_0\simeq 0.2$ K \cite{butterfly} between the two pairs 
of Kramers-degenerate levels. The aim of this paper is to 
study the decoherent 
influence of the environment upon this tunneling, i.e.
the decoherence between the states $|S=1/2,S_z=+1/2\rangle$ and
$|S=1/2,S_z=-1/2\rangle$.

First, we consider decoherence caused by the spin-lattice relaxation. 
The rate of the
relaxation due to direct one-phonon processes can be estimated 
\cite{abragam,blin} as
\begin{equation}
\label{sldir}
\left(\tau _{sl}^{dir}\right)^{-1}= 9\pi\frac{\left|V_{sl}\right|^2}{Mv^2}
 \left(\frac{\Delta_0}{\theta }\right)^3 
 \coth{\left(\frac{\Delta_0}{2T}\right)},
\end{equation}
where $\Delta_0$ is the tunneling splitting of the ground state doublets,
$T$ is the
temperature, $v\approx 2800$ m/s \cite{butterfly} is the 
sound velocity in the molecular crystal,
$M\approx 2.3\cdot 10^3$ a.m.u. is the mass 
of the molecule,
$\theta=(6\pi^2 v^3/\Omega_0)^{1/3}\simeq 70$ K is the Debye temperature
($\Omega_0$ is the volume per molecule), and $V_{sl}$ is the
characteristic ``modulation'' of spin energy under long-wavelength
acoustic deformation. At present, the physical mechanism of
spin-lattice coupling is unclear, but the value of $V_{sl}\simeq 2.6$ K
has been estimated from the available experimental data 
\cite{butterfly}. As a result, the estimate is
$\left(\tau_{sl}^{dir}\right)^{-1} \simeq 2\, T\cdot 
10^{-11}$ K (where $T$ is the temperature in Kelvins). 
Here and below, we put $\hbar = k_B=1$ and
express all quantities, including relaxation time, in the same
units (Kelvins).
Also, there is a contribution from Raman two-phonon 
processes, but at low temperatures the corresponding relaxation time 
$\tau_{sl}^R$ is very long: $\tau_{sl}^{dir}/\tau_{sl}^R\simeq
T^6/(Mv^2\Delta_0^2\theta^3)\ll 1$ \cite{abragam,blin}, and can
be neglected.

We also consider Orbach two-step relaxation via the excited 
levels $S=3/2$ \cite{blin} (see Fig.\ \ref{v15scheme}): 
\begin{equation}
\left(\tau _{sl}^{Or}\right)^{-1}=9\pi\frac{|V_{sl}|^2}{Mv^2}
  \left(\frac{\Delta_1}{\theta}\right)^3\exp{\left( 
  -\frac{\Delta_1}{T}\right)} 
\label{Orb}
\end{equation}
and for the parameters of V$_{15}$ we have 
$\left(\tau _{sl}^{Or}\right)^{-1}\simeq 10^{-8} \exp{(
  -\Delta_1/T)}$ K. Here, we assume that the spin-lattice matrix element
$V_{sl}$ is of the same order as above (about 2.6 K). 

Along with triggering Orbach processes, the excitation of molecules 
to the level
$S=3/2$ leads to a variation of the dipolar field 
exerted on a given molecule. As time goes by,
some of the excited molecules relax back to $S=1/2$,
while other molecules go up to the level $S=3/2$, and the
dipolar field at a given point in the crystal fluctuates 
with time.
In this paper, we use a mean-field approach to take into account
the dipolar fields acting on molecules; it is justified since
we are dealing with the case of relatively high temperatures
(in comparison with the energy of dipolar interactions $\Gamma_0$)
and long-range dipolar forces. Within the mean-field approach,
the dipolar field of the molecule with the spin ${\bf S}_2$ 
(equal to 3/2) can be imagined as a sum of the field created
by a spin ${\bf S}_1$ (equal to 1/2) and a field created by
the spin ${\bf S}'={\bf S}_2-{\bf S}_1$. Thus, the total
dipolar field is a sum of two fields: the static 
demagnetizing field created by a uniform medium of
spins 1/2, and a purely fluctuating field $h$ created by the spins 
${\bf S}'$. The spins ${\bf S}'$ are distributed approximately uniformly 
over the sample at any instant, and their number $N_1$ 
is small in comparison with the total number $N$ of
molecules, $N_1=N\exp{(-\Delta_1/T)}$, so the
fluctuating field $h$ at any instant obeys the Cauchy (Lorentz) 
distribution (Chapter IV, Ref.\ \cite{abragam}):
\begin{equation}
P\left( h\right) =\frac{\Gamma }{\pi }\frac{1}{h^{2}+\Gamma ^{2}}
\label{Cauchy}
\end{equation}
with $\Gamma =\Gamma_0 (N_1/N)=\Gamma_0\exp{\left( -\Delta_1/T\right)}$,
where
$\Gamma_0\simeq 10^{-4}$ K is of order of the dipole-dipole interaction 
energy in the ground state. Note that the fluctuating field $h$ is
measured against the total static field, including the static dipolar
field. A comparison with Eqs.\ (\ref{sldir}) and
(\ref{Orb})
shows that at $T>0.2$ K the distribution width $\Gamma$ is 
much larger than $1/\tau_{sl}$, so that the fluctuating field $h$
destroys coherence much faster than phonons do. Therefore,
the fluctuating dipole-dipole field constitutes an important 
decoherence factor.

To estimate the correlation time of the dipolar field fluctuations,
we note that the field changes
when excited molecules relax back to the $S=1/2$
level (and, according to the principle of detailed balance,
the same number of molecules go to the level $S=3/2$). The
transition from $S=3/2$ to $S=1/2$ proceeds via emission of phonons 
of energy $\Delta_1$. The rate of this transition is proportional to 
$\Delta_1^3$ (the number of phonons with energy $\Delta_1$), 
and can be calculated using the Fermi's golden rule 
(Chapter 10, Ref.\ \cite{blin}):
\begin{equation}
\tau_c^{-1}=9\pi\frac{|V_{sl}|^2}{Mv^2}\left( \frac{\Delta_1}
  {\theta}\right)^3,
  \label{tauc}
\end{equation}
which satisfies the condition of detailed balance between
the levels $S=3/2$ to $S=1/2$ (cf.\ Eqs.\ (\ref{tauc}) and
(\ref{Orb}) representing the rates of transitions ``up'' and
``down''), so the level populations remain constant in time.
During the time $\tau_c$, a majority of the molecules situated
initially in the state $S=3/2$ relax to $S=1/2$, and
other molecules are excited, causing the field to fluctuate.
Thus, $\tau_c$ is the correlation time for 
the fluctuating dipolar field. The estimate gives
$\tau_c^{-1} \simeq 10^{-8}$ K for V$_{15}$, so that
$\Gamma\tau_c\ll 1$ at $T<0.5$ K, i.e. the field fluctuations
are fast in comparison with their amplitude.

Now, let us consider the hyperfine fields which constitute
an important source of decoherence \cite{stamp}. 
A typical time for fluctuations of the hyperfine field $\tau_n$
is of order of the linewidth of nuclear magnetic resonance and 
can be estimated as dipole-dipole interactions between different 
nuclei \cite{abragam}: 
$1/\tau_n\sim (\mu_n/\mu_e)^2\Gamma_0$, where $\mu_n$, $\mu_e$ are
nuclear and electronic magnetic moments, respectively.
Therefore, for the temperatures $T>\Delta_1/[2\ln{
(\mu_n/\mu_e)}]\simeq 0.2$ K one has $\tau_n\Gamma \gg 1$ and 
hyperfine fields can be
considered as static for time intervals of order $\Gamma^{-1}$.
As will be shown below, $\Gamma$ defines the relaxation (decoherence)
time, so that hyperfine fields can be combined with static demagnetizing
fields to give some total static mean-field bias $h_0$ of the 
doublet levels. 
This bias is determined mainly by the hyperfine field exerted on
a molecule, which is about $\Gamma_{hf}\simeq 5\cdot 10^{-2}$ K 
\cite{butterfly} (demagnetizing
fields are weaker), and is of order of the tunneling splitting $\Delta_0$.
Therefore, for a large fraction of the molecules 
the levels $|S_z=+1/2\rangle$ and $|S_z=-1/2\rangle$ are rather close
to resonance. This is radically different from the case of
strongly anisotropic molecular magnets (such as Mn$_{12}$ or 
Fe$_8$) where the ground-state tunneling splitting is much
smaller than hyperfine fields.

Finally, we consider the static
dipolar interaction $\Gamma_0\simeq 10^{-4}$ K between the molecules 
situated in the lowest four 
states (with $S=1/2$). Longitudinal dipolar interactions 
(the terms $S^1_zS^2_z$) are 
included in the mean field along with the static hyperfine field 
$\Gamma_{hf}$, 
and can be neglected in comparison with the latter 
(since $\Gamma_0\ll \Gamma_{hf}$). The terms $S^1_zS^2_x$ etc.
within the mean-field approximation just
change the tunneling splitting negligibly (since $\Gamma_0\ll \Delta_0$).
But the flip-flop terms ($S^1_xS^2_y$ etc.) cannot be
incorporated into the mean-field scheme. Flip-flop between two
molecules is a transition from the state 
$|S_z^1=+1/2, S_z^2=-1/2\rangle$ to the state 
$|S_z^1=-1/2, S_z^2=+1/2\rangle$. The matrix element of this 
transition is of order $\Gamma_0$, but the energy difference
between the initial and final states is determined by the
difference in local mean fields acting on the two molecules,
which is of order $\Gamma_{hf}\gg\Gamma_0$. In this
situation, known as Anderson localization,
the levels of the molecule do not widen at all, and no spin diffusion
is present. The localization can be lifted due to the dynamic change 
of the hyperfine field at the molecule, but this happens on 
a timescale $t\sim \tau_n$. At temperatures $T>0.2$ K 
the coherence is already lost at these times, due to thermoactivated
dipolar field fluctuations ($\Gamma\tau_n\gg 1$).
At lower temperatures, the mean-field approach is not valid, and
the intermolecular correlations should be taken into account.

Summarizing the discussion above, the dipolar dynamic 
fluctuations constitute an important source
of decoherence at 0.2 K$<T<$ 0.5 K.
Let us formulate now a model for magnetic relaxation under
the fluctuating dipolar field ${\bf h}= h_x{\bf e}_x +h_y{\bf e}_y 
+h_z{\bf e}_z$. We consider a two-level system (the
levels $S_z=\pm 1/2$ for V$_{15}$) with the static
tunneling splitting $\Delta_0$ 
and static mean-field bias $h_0$ (the latter is governed mainly 
by the hyperfine static fields, since the demagnetizing fields
are much weaker).
The system is described by the density matrix $\rho$ written
in the basis formed by the levels $S_z=\pm 1/2$. 
Its equation of motion is
\begin{equation}
\dot\rho = i\left[ \rho, {\cal H}\right]
\end{equation}
where ${\cal H} = -(\Delta_0+h_x)\sigma_x -h_y\sigma_y 
-(h_0+h_z)\sigma_z$ is the Hamiltonian of the system ($\sigma_{x,y,z}$
are the Pauli's matrices). It can be conveniently
written as
\begin{eqnarray}
\label{ro}
\dot x &=& -h_y y - (\Delta_0+h_x)z \\
 \nonumber
\dot y &=& h_y x + (h_0+h_z)z \\ 
 \nonumber
\dot z &=& (\Delta_0+h_x)x-(h_0+h_z)y 
\end{eqnarray}
by introducing the variables: $x=(\rho_{11}-\rho_{22})/2$,
$y=(\rho_{12}+\rho_{21})/2$ and $z=(\rho_{12}-\rho_{21})/(2i)$.
The static fields $\Delta_0$ and $h_0$ can be eliminated by two
rotations of the co-ordinate frame:
\begin{eqnarray}
\label{trans}
x &=& X\cos{\varphi} - (Y\cos{Et}+Z\sin{Et})\sin{\varphi}\\
 \nonumber
y &=& X\sin{\varphi} + (Y\cos{Et}+Z\sin{Et})\cos{\varphi}\\
 \nonumber
z &=& -Y\sin{Et}+Z\cos{Et},
\end{eqnarray}
where $\sin{\varphi} =\Delta_0/E$, $\cos{\varphi}=h_0/E$, 
$E=\sqrt{\Delta_0^2+h_0^2}$, and Eqs.\ (\ref{ro}) take the form 
\begin{eqnarray}
\label{ro2}
\sqrt{2}\,\dot X &=& \left(h_{2a}-h_{3b}\right) Y
  -\left(h_{2b}+h_{3a}\right) Z \\
  \nonumber
\sqrt{2}\,\dot Y &=& \sqrt{2}\,h_1 Z + \left(h_{3b}-h_{2a}\right) X \\
  \nonumber
\sqrt{2}\,\dot Z &=& \left(h_{2b}+h_{3a}\right)X - \sqrt{2}\,h_1 Y.
\end{eqnarray}
The random fields acting on the system are
$h_1 = h_z\cos{\varphi} + h_x\sin{\varphi}$, $h_{2a,3a} = 
\sqrt{2}\, h_{2,3}\sin{Et}$, and $h_{2b,3b} = \sqrt{2}\, h_{2,3}\cos{Et}$, 
where $h_2= -h_z\sin{\varphi} + h_x\cos{\varphi}$ and $h_3=h_y$.
As we discussed above, $h_{x,y,z}$ are independent 
random fields, at any instant
distributed with the same law (\ref{Cauchy}); the same is true 
for $h_{1,2,3}$. 
Since $E\tau_c\geq\Delta_0\tau_c \gg 1$, one can consider 
$h_{2a,3a}$ and $h_{2b,3b}$ as independent, and in Eq.\ (\ref{ro2})
we have several independent fluctuating fields with the same 
Cauchy distribution and with very short autocorrelation time $\tau_c$.

Eqs.\ (\ref{ro2}) can be imagined as describing the evolution of 
a system with the Hamiltonian $H=H_1+H_2+H_3$
\begin{equation}
\label{rororo}
\dot{\bf R} = -iH{\bf R}
\end{equation}
where ${\bf R}=(X,Y,Z)$, $H_1=\sqrt{2}\,(h_{2a}-h_{3b}) S_1$, 
$H_2=-\sqrt{2}\,(h_{2b}+h_{3a}) S_2$, $H_3=h_1 S_3$,
and the noncommuting matrices $S_{1,2,3}$ are
\begin{equation}
S_1=\left(
\begin{array}{ccc}
0 & -i & 0 \\ 
i & 0 & 0 \\ 
0 & 0 & 0
\end{array}
\right),
 \,
S_2=\left(
\begin{array}{ccc}
0 & 0 & -i \\ 
0 & 0 & 0 \\ 
i & 0 & 0
\end{array}
\right),
 \,
S_3=\left(
\begin{array}{ccc}
0 & 0 & 0 \\ 
0 & 0 & -i \\ 
0 & i & 0
\end{array}
\right).
\end{equation}
The formal solution of Eqs.\ (\ref{rororo}) can be represented in a 
path-integral-like form, by splitting the time interval $(0,t)$ into 
$N\gg 1$ equal pieces of length $\epsilon=t/N$:
\begin{equation}
\label{sol}
{\bf R}(t) = \exp{[-i\epsilon H(t_{N-1})]} \dots 
  \exp{[-i\epsilon H(0)]} {\bf R}(0) 
\end{equation}
where $t_n=n\epsilon$. Each of the matrices $H$ is proportional to
the fluctuating fields $h_{1,2,3}$, so if we choose $\epsilon\ll\Gamma^{-1}$
the Trotter decomposition formula \cite{trotter} can be applied 
to each factor: 
\begin{equation}
\exp{\left(-i\epsilon\sum H_k\right)} =\prod \exp{(-i\epsilon H_k) +
  {\cal O}(\epsilon^2)},
\label{trot}
\end{equation}
where $k=1,2,3$. The correlation time of all the fields $h_{1,2,3}$ is
$\tau_c$, so $H_k(t)$ and $H_k(t+\epsilon)$ in Eq.\ (\ref{trot}) are 
decorrelated if $\epsilon\gg\tau_c$. Choosing 
$\tau_c \ll\epsilon\ll \Gamma^{-1}$, 
each term in the products (\ref{sol}) and (\ref{trot}) can be
averaged independently over different realizations of the
random processes represented by the fields $h_{1,2,3}$
thus giving the answer: 
\begin{eqnarray}
\label{av}
\langle X(t)\rangle &=& X(0)\exp{(-2\sqrt{2}\Gamma t)} \\
\nonumber
\langle Y(t)\rangle &=& Y(0)\exp{[-(\sqrt{2}+1)\Gamma t]} \\
\nonumber 
\langle Z(t)\rangle &=& Z(0)\exp{[-(\sqrt{2}+1)\Gamma t]}
\end{eqnarray}
These results, together with Eq.(\ref{trans}), represent an exact 
solution of the problem.
The situation considered here is similar to that found in spin
resonance, and the results can be conveniently expressed in 
corresponding terms. The dynamics of the density matrix elements
is represented as a sum of two terms: damped oscillations with 
the frequency $E$ (with transverse relaxation rate 
$T^{-1}_2=(\sqrt{2}+1)\Gamma$) and pure damping (with longitudinal
damping rate $T^{-1}_1=2\sqrt{2}\Gamma$). 
The decoherence times $T_{1,2}$ both are of order $\Gamma^{-1}$. 
This holds in spite 
of the smallness of $\tau_c$, due to the peculiar properties of 
the Cauchy distribution:
for Gaussian fluctuations with variance $\sigma^2$ we would have 
much smaller relaxation rate $\sigma^2\tau_c$ (motional narrowing
\cite{abragam}). On the other hand, if $\tau_c$ were very
large then the dipolar field would be almost static, and 
the decoherence time for a single molecule would be determined by
hyperfine fields, as it is for Mn$_{12}$ or Fe$_8$ \cite{stamp}.

Nevertheless, for V$_{15}$ the decoherence rate is still small
enough: $\Gamma/\Delta_0\simeq 2\cdot 10^{-7}$ at $T=0.5$ K,
i.e. the system tunnels about 
5,000,000 times before the tunneling oscillations are wiped out by
decoherence. We emphasize that each tunneling in V$_{15}$
is {\it not\/} a single-spin event: it takes place between the two states
of the whole molecule. It is a tunneling of an antiferromagnetic
system with small uncompensated spin, and {\it all\/} 15 spins 
are involved. 

Summarizing, we considered possible sources of decoherence in
V$_{15}$ molecules between the states 
$S_z=\pm 1/2$ of ground-state doublets. We found that in the 
temperature region
0.2--0.5 K the main source of decoherence is the fluctuating
dipolar field created by the molecules, which are thermally
activated to the higher $S=3/2$ level. Based on an exactly solvable
model, a rather low decoherence rate is found: 
about 5,000,000 tunneling events occur before the coherence
is destroyed. Such a low decoherence rate is unusual for
magnetic systems of mesoscopic size at these temperatures.

Authors would like to thank W.\ Wernsdorfer, I.\ Chiorescu and B.\ Barbara  
for helpful discussions. 
This work was partially carried out at the Ames Laboratory, which 
is operated for the U.\ S.\ Department of Energy by Iowa State 
University under Contract No.\ W-7405-82 and was supported by 
the Director for Energy Research, Office of Basic Energy Sciences 
of the U.\ S.\ Department of Energy. This work was partially supported 
by Russian Foundation for Basic Research, grant 98-02-16219.

\begin{figure}
\caption{Sketch of the V$_{15}$ molecule. The spins $B$ and $B'$ 
of the upper hexagon form a very stiff antiferromagnetic
dimer: $J_{BB'}\sim 800$ K. The spin $A$ from the central triangle
is coupled antiferromagnetically
to both $B$ and $B'$, so $J_{AB}$ and $J_{AB'}$ cancel each other 
to a large extent ($J_{AB}, J_{AB'}\ll J_{BB'}$). 
This frustration effectively decouples the triangle spins 
from the hexagons. Due to $C_{3v}$ symmetry of the molecule, the
situation is the same for all the triangle spins.}
\label{v15scheme}
\end{figure}

\end{document}